# Comparative Analysis of Lightweight Kubernetes Distributions for Edge Computing: Security, Resilience and Maintainability

Diyaz Yakubov[0009-0000-7989-4614] and David Hästbacka[0000-0001-8442-1248]

Tampere University, Tampere FI-33014, Finland
{diyaz.yakubov, david.hastbacka}@tuni.fi

**Abstract.** The increasing demand for real-time data processing in Internet of Things (IoT) devices has elevated the importance of edge computing, necessitating efficient and secure deployment of applications on resource-constrained devices. Kubernetes and its lightweight distributions — k0s, k3s, KubeEdge, and OpenYurt — extend container orchestration to edge environments, but their security, reliability, and maintainability have not been comprehensively analyzed. This study compares Kubernetes and these lightweight distributions by evaluating security compliance using kube-bench, simulating network outages to assess resiliency, and documenting maintainability. Results indicate that while k3s and k0s offer superior ease of development due to their simplicity, they have lower security compliance compared to Kubernetes, KubeEdge, and OpenYurt. Kubernetes provides a balanced approach but may be resource-intensive for edge deployments. KubeEdge and OpenYurt enhance security features and reliability under network outages but increase complexity and resource consumption. The findings highlight trade-offs between performance, security, resiliency, and maintainability, offering insights for practitioners deploying Kubernetes in edge environments.

## 1 Introduction

The rapid spreading of Internet of Things (IoT) devices and the increasing demand for real-time data processing have propelled edge computing to the forefront [4, 8, 9]. By bringing computation and data storage closer to the data sources, edge computing reduces latency, conserves bandwidth, and enables faster decision-making. However, deploying applications at the network edge introduces challenges, particularly concerning security, resilience, and maintainability.

Edge computing environments are often resource-constrained, and the devices are frequently distributed across vast geographical areas, including intermittent network connectivity. Such conditions amplify security vulnerabilities due to increased attack surfaces and physical accessibility. Resilience becomes a concern as network disruptions can lead to service outages, and maintaining operations across distributed nodes adds complexity to system management.

Container orchestration platforms like Kubernetes (k8s) have revolutionized application deployment and management in cloud environments through automation and scalability. Recognizing the potential benefits for edge computing, several lightweight kubernetes distributions (later KD), e.g., k0s, k3s, KubeEdge, and OpenYurt, have been developed to extend k8s functionalities on the edge.

While performance optimization has been a central focus in prior studies comparing lightweight KDs [3, 6, 7, 10], there is a research gap on security, resiliency, and maintainability. This study provides a comparative analysis of selected lightweight KDs focusing on the following research questions: **RQ1:** How do lightweight KDs differ in terms of security compliance, and what are the implications for edge computing applications? **RQ2:** How do these distributions perform under simulated network outage scenarios, and what insights does this provide about their resiliency? **RQ3:** What are the maintainability challenges associated with each distribution when deployed in edge environments?

This study highlights trade-offs between security, resilience and maintainability of different distributions, providing insight to practitioners and researchers aiming to deploy kubernetes in resource-constrained edge environments. The paper concludes with recommendations for selecting the distribution based on specific needs, balancing security, resiliency, and maintainability.

## 2 Background and Related Work

### 2.1 Lightweight Kubernetes Distributions for Edge Computing

Kubernetes (k8s)[1] has become the de facto standard for container orchestration, automating deployment, scaling, and management of containerized applications. Various distributions have been developed for different use cases, also focusing on resource-constrained environments such as edge devices and IoT gateways. k3s[2] is a lightweight KD developed by Rancher Labs, now part of SUSE. Designed to have a smaller footprint, k3s is packaged as a single binary (around 100 MB) and requires less memory and CPU than the standard Kubernetes. k0s[3] is another lightweight KD, created by Mirantis. Similar to k3s, k0s aims to provide a streamlined and easy-to-install Kubernetes experience with minimal resource consumption. It combines all necessary components into a single binary and supports various storage options, including etcd and SQLite. OpenYurt[4], by Alibaba Cloud, extends Kubernetes for the edge while retaining standard APIs, and enhancing edge autonomy and cloud-edge synergy. KubeEdge[5], a CNCF

1 https://kubernetes.io/
2 https://k3s.io/
3 https://k0sproject.io/
4 https://openyurt.io/
5 https://kubeedge.io/

project, brings Kubernetes to edge environments by providing infrastructure and APIs to manage applications on edge nodes as if they were in the cloud.

### 2.2 Related Work

Koziolek et al. [3] conducted a detailed performance comparison of lightweight KDs, including MicroK8s, k3s, k0s, and MicroShift. They focused on resource usage and control plane and data plane performance under stress scenarios. However, they did not address security or reliability aspects.

Cilic et al. [10] evaluated the applicability of Kubernetes, k3s, KubeEdge, and ioFog in edge environments. They analysed deployment complexity, memory footprint, and service startup times. Although they highlighted challenges specific to edge environments, such as resource constraints and service management, their study did not delve deeply into security compliance or maintainability.

Kotopulis Ostinelli et al. [5] focused on performance evaluation and emulation of LTE/5G networks over a lightweight open platform using k3s. They demonstrated the using of lightweight KDs in telecommunications environments, emphasizing scalability, automation, and reliability in 5G deployment scenarios.

Fogli et al. [2] evaluated the performance of k8s, k3s, and KubeEdge in adaptive and federated cloud infrastructures, particularly in disadvantaged tactical networks with limited bandwidth and high latency. They found KubeEdge's performance superior in maintaining cluster stability under degraded network conditions, emphasizing the significance of reliability in such environments.

## 3 Methodology

**Experimental Setup** Test cluster: *Master Node:* Intel NUC (i7-10710U/64GB DDR4/1TB NVMe) device serving as the control plane. *Worker Nodes:* Three Raspberry Pi 4 Model B (4GB) devices acting as resource-constrained edge nodes. *Auxiliary Node:* Intel NUC (i7-10710U/64GB DDR4/1TB NVMe) for data collection and storage. To minimize variables and focus on the KDs themselves, all devices were connected via LAN and configured with the same Ubuntu 22.04.2 making a closed system. More details are in the project's repository[6]. The KDs were installed and configured according to their official documentation.

**Security Assessment (RQ1)** For the security evaluation, we ran *kube-bench* on each distribution to assess compliance with the CIS Kubernetes Benchmark[7]. We developed a Security Score to quantify the compliance level, assigning criticality weights to each check and applying coefficients based on the test outcomes (PASS, WARN, FAIL).

[6] https://github.com/DiyazY/iot-edge/blob/main/src/diagrams/deployment_diagram-full-v3-colored.pdf

[7] https://github.com/aquasecurity/kube-bench

**Security Score**: Calculated based on the weighted results from *kube-bench*, reflecting compliance with critical security benchmarks. To each security check a criticality weight[8] from 1 (Informational) to 5 (Critical) was assigned, reflecting the potential impact on the cluster's security. The outcome of each check was assigned a coefficient: PASS (1), WARN (0.5), and FAIL (-1). The Security Score for each distribution was calculated using the formula: Security Score = $(Sum(CriticalityWeight * ResultCoefficient))/MaximumPossibleScore * 100\%$

**Resiliency Assessment under Network Outages (RQ2)** To simulate network outages and evaluate resiliency, we conducted two types of tests: *1) Master Node Outage*: Temporarily disconnected the master node from the network for a fixed duration while workloads were running, then observed the cluster's behavior and recovery process. *2) Worker Node Outage*: Disconnected a worker node under similar conditions and monitored the impact on workloads and the node's reintegration into the cluster. *k-bench*[9] was used to generate workloads.

We used the following metrics to evaluate: *Recovery Time* - Time taken for the cluster to return to a fully operational state after a simulated outage. *Resource Utilization* - CPU, memory, network bandwidth, and disk I/O during outages and recovery phases. *netdata*[10] was used to monitor resource utilization. To ensure statistical significance, each test was repeated five times, and the results were averaged. Between tests, we allowed sufficient cool-down periods to stabilize the system and avoid carry-over effects. The test flow diagram and other tests results (idle, light and heavy load tests, long Network outage test) can be found in the project's repository[11].

**Maintainability Evaluation (RQ3)** was assessed by documenting the following: *Installation Complexity*: Recorded the time and steps required to install each distribution, noting any challenges encountered. *Configuration and Updates*: Evaluated the ease of configuring cluster components and performing updates or patches. *Operational Overhead* : Monitored the need for manual interventions during tests, such as node reboots or troubleshooting procedures.

The metrics include: *Setup Time* - Duration and complexity of initial installation. *Update Complexity* - Steps and time to apply updates or patches. *Manual Interventions* - Frequency and nature of required manual maintenance tasks.

8 https://github.com/DiyazY/iot-edge/blob/main/src/kube-bench-results/final-reports/SECURITY.md
9 https://github.com/vmware-tanzu/k-bench
10 https://www.netdata.cloud/
11 https://github.com/DiyazY/iot-edge/blob/main/src/diagrams/test_flow-2nd.pdf

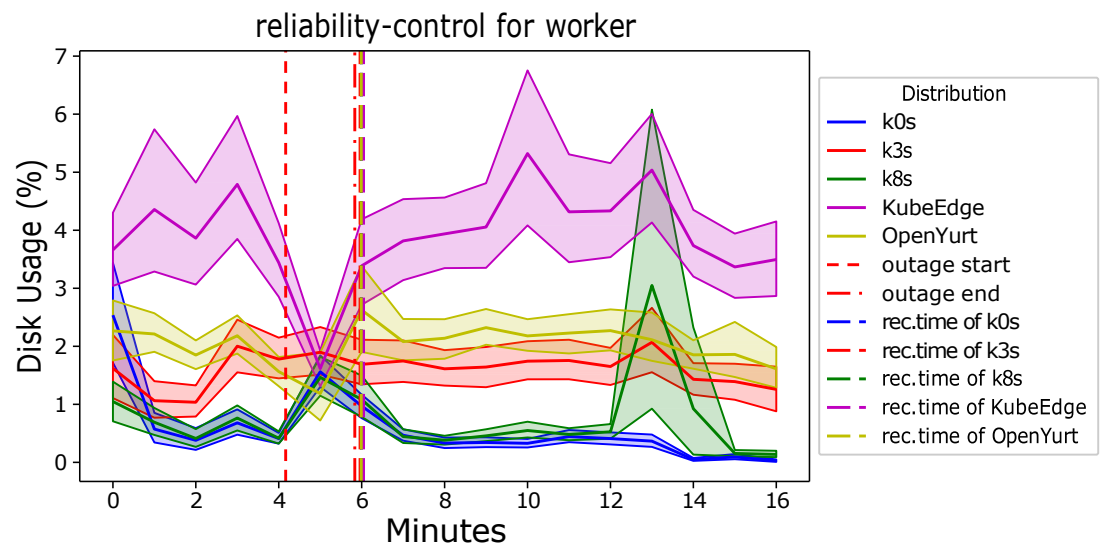

**Fig. 1.** Disk I/O usage on worker nodes during master node outage

# 4 Results

## 4.1 Security Assessment (RQ1)

Using the metrics defined in section 3 the Security Scores for the evaluated distributions are as follows: k0s scored 23.69%, k3s scored 7.21%, and k8s, KubeEdge, and OpenYurt each scored 55.00%.

The results indicate that KDs based on the standard platform — such as k8s, KubeEdge, and OpenYurt — benefit from comprehensive default security configurations, leading to higher security compliance scores. In contrast, k0s and k3s, designed with a minimalist approach to optimize resource efficiency by omitting certain security features, achieve lower scores. Consequently, these lightweight distributions may require additional configuration and hardening to meet security best practices, especially in edge environments.

## 4.2 Resiliency Assessment (RQ2)

**Master Node Outage**, during the *reliability_control* test, the master node was disconnected for 100 seconds while workloads were running. Observations include: *Resource Utilization*: The failed master node exhibited increased network, CPU, and disk I/O usage, but decreased RAM usage. *Worker Nodes*: Network utilization increased, and disk I/O varied among distributions. *Disk I/O Behavior*: KubeEdge and OpenYurt decreased disk I/O usage on worker nodes during the outage, while k0s, k3s, and k8s showed increased disk I/O (Figure 1).

**Worker Node Outage**, in the *reliability_worker* test, a random worker node was disconnected for 100 seconds. Key findings include: *Failed Worker Node*: Showed increased disk I/O and RAM utilization during the outage. *KubeEdge Behavior*: Demonstrated prolonged high disk I/O utilization after recovery and a delayed return to normal network usage compared to other distributions. *Network Utilization*: KubeEdge's network usage decreased during the outage and gradually increased upon recovery, suggesting message preservation and retransmission mechanisms. Figure 2 depicts disk I/O and Network usages in this case.

Additionally, KubeEdge exhibited a delayed completion of the *reliability_worker* test compared to other distributions, potentially due to its edge-specific features.

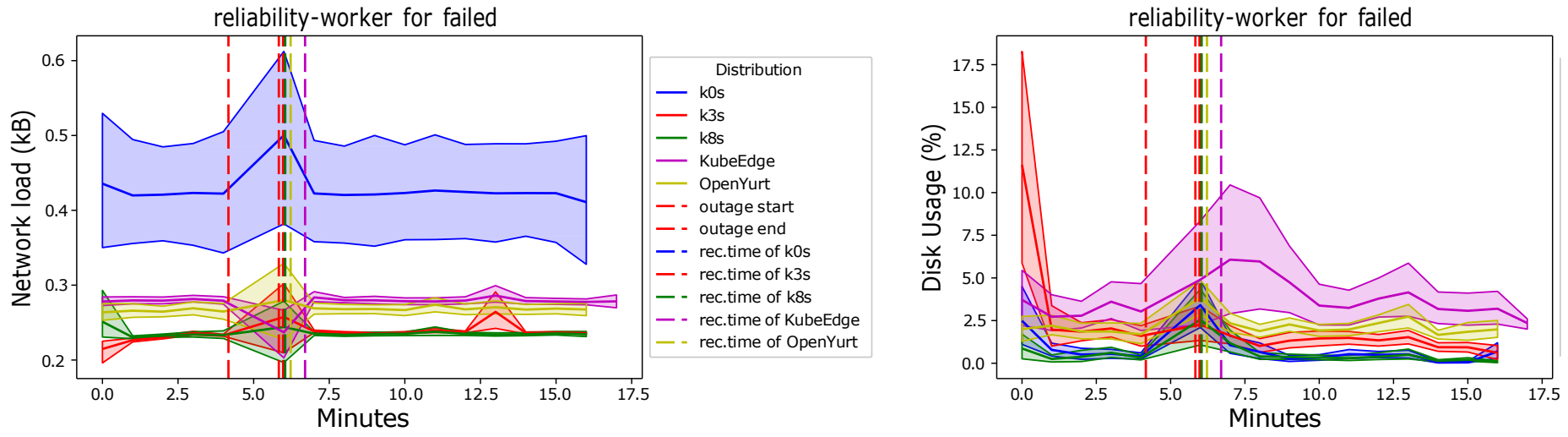

**Fig. 2.** Disk I/O (left) and Network (right) usages on failed worker node during outage

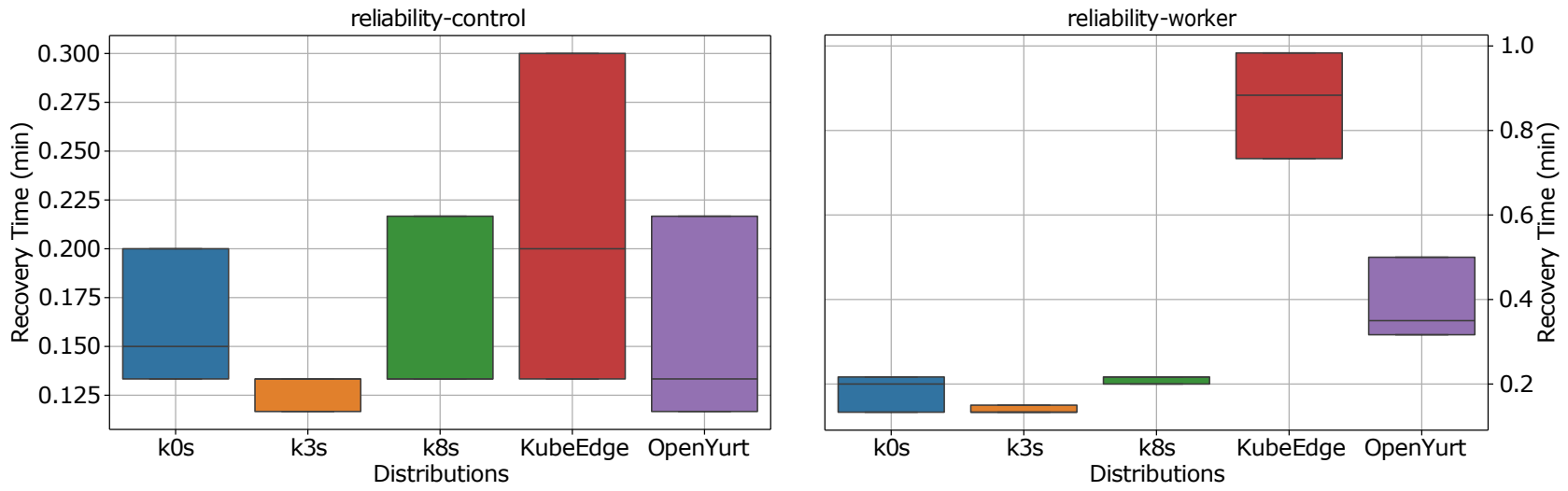

**Fig. 3.** Recovery Time: Master node unavailable (right), Worker node unavailable (left)

If we examine the recovery times in Figure 3, it is noticeable that KubeEdge and OpenYurt have prolonged recovery times, which is particularly obvious in worker node outage scenarios since the workloads are running on worker nodes.

**Analysis of Resiliency Findings**, the observations suggest that KubeEdge and OpenYurt implement mechanisms to preserve messages during network outages, enhancing resiliency at the cost of increased resource consumption and longer recovery times. In contrast, k0s, k3s, and k8s show faster recovery and lower resource usage but may not preserve in-transit messages during outages.

### 4.3 Maintainability Assessment (RQ3)

**Analysis of Maintainability Findings** Our maintainability assessment indicates that k0s and k3s offer superior ease of use and lower maintenance efforts, making them suitable for edge environments. In contrast, KubeEdge and OpenYurt, while providing advanced features for edge computing, introduce additional complexity that may challenge maintainability.

Tables 1 and 2 summarize the maintainability aspects of each distribution. Consistent with our observations, [1] emphasizes that differences in automation capabilities and day-2 operations management significantly impact the maintainability of KDs. These findings support the notion that while advanced features enhance functionality, they may negatively impact maintainability.

## 5 Discussion

While k3s and k0s demonstrate superior maintainability due to simplicity and ease of setup, they exhibit lower security compliance compared to k8s, KubeEdge, and OpenYurt. k8s and its edge-optimized variants provide better security by default but introduce extra complexity in deployment and maintenance. In terms of network outage resiliency, KubeEdge and OpenYurt exhibit unique behaviors aimed at preserving system state and ensuring message durability. This is evident in their resource utilization patterns during simulated outages, where they consume more disk I/O and exhibit delayed recovery times. These mechanisms enhance resiliency in intermittent network conditions common in edge environments but may impact performance and resource consumption. Maintainability assessments reveal that minimalist designs of k3s and k0s reduce operational overhead. In contrast, the added complexity of KubeEdge and OpenYurt necessitates more extensive maintenance.

## 6 Conclusion

This study analysed lightweight KDs, namely k0s, k3s, k8s, KubeEdge and OpenYurt, with a focus on security, resilience, and maintainability in edge computing environments. Our findings indicate that k3s and k0s excel in maintainability due to their simplicity and ease of deployment, making them suitable for resource-constrained edge settings where operational overhead must be minimized. However, they exhibit lower security compliance, highlighting the need for additional security configurations to meet best practices. KubeEdge and OpenYurt offer features tailored for edge computing and demonstrate higher security compliance and mechanisms to enhance resiliency under network outages. These benefits come at the cost of increased complexity and resource consumption, impacting maintainability and performance. The standard Kubernetes distribution (k8s) provides a balanced approach but may still be resource-intensive for some edge deployments. Selecting the appropriate distribution depends on the specific requirements and constraints of the deployment scenario, including the priority of security features, tolerance for complexity, and available resources for maintenance. Practitioners must carefully consider the trade-offs between security, resiliency, and maintainability to select a distribution that aligns with their operational needs. Future work on KDs should balance security and resource constraints without compromising maintainability, enhance automation to reduce operational overhead, explore adaptive resiliency mechanisms that balance performance and resource utilization under varying network conditions, standardize edge features to improve interoperability and ease of use, and finally, employ AI-driven agents for predictive maintenance. Likewise, future test suites should cover various outage and network saturation scenarios at a larger scale.

**Table 1.** Maintainability Comparison of Kubernetes Distributions

| Distribution | Installation Effort (hours) | Update Process | Operational Overhead |
|---|---|---|---|
| k0s | Easy (3h) | Simple, manual/automated | Low |
| k3s | Easy (2h) | Simple, manual/automated | Low |
| k8s | Moderate (7h) | Complex, manual | Moderate |
| KubeEdge | Difficult (14h) | Complex, manual | High |
| OpenYurt | Difficult (14h) | Complex, manual | High |

**Table 2.** Maintainability Evaluation of Kubernetes Distributions

| Installation and Setup | Configuration and Updates | Operational Overhead | Documentation and Community Support |
|---|---|---|---|
| *k0s and k3s*: Offered the simplest installation processes, with minimal steps and straightforward commands. Single-binary designs simplify deployment. *k8s*: Required more complex setup due to its modular components and configuration options. *KubeEdge and OpenYurt*: KubeEdge required separate installation of cloud and edge components, along with configuration of mesh communication network for node-to-node communication. OpenYurt needed to set up its edge autonomy features carefully. | *k0s and k3s*: Provided ease of configuration with sensible defaults. Updating these distributions was relatively straightforward, often involving simple commands or replacing the binary. *k8s*: Updates and configuration changes required careful planning to avoid disruptions, given its complexity and the potential impact on running services. *KubeEdge and OpenYurt*: Updates were more involved due to the additional components and dependencies. Configuration changes had to account for edge-specific features and potential connectivity issues. | *k0s and k3s*: Demonstrated high maintainability with minimal need for manual interventions during tests. Their resilience to restarts and system reboots reduced operational overhead. *k8s*: Required moderate maintenance efforts. While stable, the complexity of its components occasionally necessitated manual adjustments. *KubeEdge and OpenYurt*: Incurred higher operational overhead. For example, OpenYurt experienced networking issues after several test cycles, necessitating restarts and troubleshooting. KubeEdge's additional features increased the maintenance burden. | *k0s and k3s*: Benefited from active communities and comprehensive documentation, aiding in troubleshooting and maintenance tasks. *k8s*: As the standard KD, it has extensive resources and community support. *KubeEdge and OpenYurt*: While documentation exists, the specialized nature of these distributions means that community support is less extensive, potentially complicating maintenance. Moreover, many pages in the official documentation are marked as TBD (To Be Defined). |

## Acknowledgement

This work is supported by funding from Business Finland in project Industry X.